\documentclass[aps,prl,preprint,nopacs,superscriptaddress]{revtex4}

\usepackage{graphicx}
\usepackage{verbatim}
\usepackage{mathrsfs}
\usepackage{gensymb}
\usepackage{ulem}

\usepackage{amsmath,amsfonts,amssymb}
\usepackage{graphicx}

\def\3{2.8in}    
\def\2{2.5in}
\def\4{3.0in}\def \beq {\begin{equation}}
\def \eeq {\end{equation}}
\begin{document}

\title{ Direct Transition Resonance in Atomically Uniform Topological Sb(111) Thin Films }

\author{Guang~Bian}
\affiliation {Laboratory for Topological Quantum Matter and Spectroscopy (B7), Department of Physics, Princeton University, Princeton, New Jersey 08544, USA}
\affiliation {Department of Physics, University of Illinois at Urbana-Champaign, 1110 West Green Street, Urbana, Illinois 61801-3080, USA}

\author{Caizhi~Xu}
\affiliation {Department of Physics, University of Illinois at Urbana-Champaign, 1110 West Green Street, Urbana, Illinois 61801-3080, USA}
\affiliation {Frederick Seitz Materials Research Laboratory, University of Illinois at Urbana-Champaign, 104 South Goodwin Avenue, Urbana, Illinois 61801-2902, USA}

\author{Tay-Rong~Chang}
\affiliation {Laboratory for Topological Quantum Matter and Spectroscopy (B7), Department of Physics, Princeton University, Princeton, New Jersey 08544, USA}
\affiliation {Department of Physics, National Tsing Hua University, Hsinchu 30013, Taiwan}

\author{Xiaoxiong~Wang}
\affiliation {College of Science, Nanjing University of Science and Technology, Nanjing 210094, China}

\author{Saavanth~Velury}
\affiliation {Department of Physics, University of California, Berkeley, California 94720, USA}

\author{Jie~Ren}
\affiliation {Department of Physics, University of Illinois at Urbana-Champaign, 1110 West Green Street, Urbana, Illinois 61801-3080, USA}
\affiliation {State Key Laboratory of Solidification Processing, School of Materials Science and Engineering, Northwestern Polytechnical University, Xi'an 710072, China}

\author{Hao~Zheng}
\affiliation {Laboratory for Topological Quantum Matter and Spectroscopy (B7), Department of Physics, Princeton University, Princeton, New Jersey 08544, USA}

\author{Horng-Tay~Jeng}
\affiliation {Department of Physics, National Tsing Hua University, Hsinchu 30013, Taiwan}
\affiliation {Institute of Physics, Academia Sinica, Taipei 11529, Taiwan}

\author{T.~Miller}
\affiliation {Department of Physics, University of Illinois at Urbana-Champaign, 1110 West Green Street, Urbana, Illinois 61801-3080, USA}
\affiliation {Frederick Seitz Materials Research Laboratory, University of Illinois at Urbana-Champaign, 104 South Goodwin Avenue, Urbana, Illinois 61801-2902, USA}

\author{M.~Zahid~Hasan}
\affiliation {Laboratory for Topological Quantum Matter and Spectroscopy (B7), Department of Physics, Princeton University, Princeton, New Jersey 08544, USA}

\author{T.-C.~Chiang}
\affiliation {Department of Physics, University of Illinois at Urbana-Champaign, 1110 West Green Street, Urbana, Illinois 61801-3080, USA}
\affiliation {Frederick Seitz Materials Research Laboratory, University of Illinois at Urbana-Champaign, 104 South Goodwin Avenue, Urbana, Illinois 61801-2902, USA}
\affiliation {Department of Physics, National Tsing Hua University, Hsinchu 30013, Taiwan}


\date{\today}

\begin{abstract}
 Atomically uniform Sb(111) films are fabricated by the method of molecular beam epitaxy on an optimized Si(111) surface. Two dimensional quantum well states and topological surface states in these films are well resolved as measured by angle-resolved photoemission spectroscopy. We observe an evolution of direct transition resonances by varying the excitation photon energy (and thus the perpendicular crystal momentum). The experimental results are reproduced in a comprehensive model calculation taking into account first-principles calculated initial states and time-reversed low-energy-electron-diffraction final states in the photoexcitation process. The resonant behavior illustrates that the topological surface states and the quantum well states are analytically connected in momentum space in all three dimensions.  

\end{abstract}

\maketitle

With the advent of nano-scale electronics, it becomes ever more important to understand the electronic structure of functional materials with reduced dimensions including quantum dots, nano wires and ultrathin films. Particularly, thin films with thickness down to tens or a few atomic layers exhibit many unusual phenomena such as electrical-resistance anomaly \cite{Ehrlich}, oscillation in superconductivity transition temperature \cite{Xue} and emergent topological order \cite{Liu}.  These quantum phenomena, caused by electron confinement, offer great opportunities for nanoscale engineering and tailoring of material properties \cite{TC}.  Another advantage of thin film systems is the reduction of bulk carriers, making the contribution from surface states more pronounced. This is important in device applications when utilizing the unusual surface (or interface) modes, like the Rashba surface states on heavy metal surfaces \cite{LaShell, Bihlmayer, Schouteden,  Ast, Bian_Bi, Hirahara, Hofmann} or the Dirac surface states in topological insulators \cite{Hasan, Zhang}. It is also important that the films be prepared with atomic-scale smoothness and a precisely known thickness. This requirement imposes stringent constraints on experimental conditions. Several key factors such as physical and chemical properties of the substrate, chemical flux rates and ratio, growth temperature, and annealing procedure determine critically the smoothness of the grown thin film, but they are highly material-dependent. So far only a few ultrathin films have been grown experimentally with high structural quality  \cite{TC, Bian_Bi, Hasegawa, Yang, Shih, ZhangYi, LiuYang, MoS2}. The electronic structure of such films can be straightforwardly mapped out by angle-resolved photoemission spectroscopy (ARPES). Even though ARPES has been widely used to study the band structure of materials, a comprehensive theoretical modeling of the photoexcitation process remains a challenge with profound impacts on both fundamental sciences and practical applications.

One thin film system of special interest is the semimetallic Sb(111) thin films with topologically nontrivial surface states. Our previous work has demonstrated the growth of Sb(111) films on Si substrate with a Bi buffer layer \cite{Bian_Sb}. However, the fuzzy photoemission spectra imply a suboptimal structural quality. In this work, we report a novel route to preparing Sb thin films with high reliability. The thickness-dependent quantum well states and topological surface states are clearly resolved by ARPES, suggesting atomic-scale smoothness of the film. Upon varying the incident photon energy, the ARPES spectra show direct-transition resonances for all states, including the topological surface states; the similarity of the resonant behavior indicates an innate relationship between the surface states and the quantum well states. The results are well explained by a model calculation based on first-principles initial states and time-reversed low-energy-electron-diffraction (TRLEED) final states. The results provide a comprehensive understanding of the photoexcitation process and illustrate the analytic connection between the topological surface states and bulk sates.

The ARPES measurements were performed at the Synchrotron Radiation Center, University of Wisconsin-Madison. A Scienta analyzer equipped with a two-dimensional detector was employed for data collection. The energy and momentum resolutions were 15 meV and 0.01 $\text{\AA}^{-1}$, respectively. First-principles calculations of the electronic structure were performed using HGH-type pseudopotentials \cite{HGH} and a plane-wave basis set. The main program employed was developed by the ABINIT group \cite{abinit}. Spin-orbit coupling was included using the relativistic LDA approximation.

The Sb thin film samples were fabricated $in~situ$ by molecular-beam-epitaxy (MBE) deposition on a properly treated Si(111) surface as schematically illustrated in Fig.~1(a). An n-type Si(111) wafer (Sb-doped with a resistivity of $\sim$0.01 $\Omega\cdot$cm) was cleaned by direct current heating to yield a 7$\times$7 reconstructed surface. It was cooled to 60 K, and about 6 $\textrm{\AA}$ of Bi was deposited on top. The sample was then annealed at 600 K for 10 minutes to yield a well ordered Bi-$\sqrt{3}\times\sqrt{3}$ surface reconstruction. We emphasize that this step was the key to the growth of smooth Sb films through our experimentation. Bi adatoms passivate the dangling bonds on the Si-7$\times$7 surface and flattened the corrugated surface. Without this surface treatment, the Sb films grown directly on the Si-7$\times$7 surface are generally so rough that the quantum well states cannot be clearly resolved by ARPES. The deposition of Sb was performed with the substrate at 60 K with the deposition rate monitored by a quartz thickness monitor. The resulting structure was annealed at 500 K to yield a smooth Sb(111) film. A sharp 1$\times$1 RHEED pattern appeared after annealing, indicative of a well ordered film structure. The crystal structure of the (111)-oriented Sb films is shown in Figs.~1(b, c). It consists of a stack of bilayers (BLs), each of which resembles a buckled graphene sheet. The bulk and (111)-projected surface Brillouin zones are shown in Fig.~1(d). The surface Brillouin zone is a simple hexagon. 

ARPES mapping along the $\bar{M}-\bar{\Gamma}-\bar{M}$ direction of the band structure of a 20 BL Sb film is shown in Fig.~2(a). Around $\bar{\Gamma}$, we observe a pair of Rahba-like surface bands crossing the Fermi level and a dense stack of electron-like quantum well bulk bands at about 0.25 eV below the Rashba surface states. At the zone boundary $\bar{M}$, there is a single quantum well conduction band right below the Fermi level accompanied by several quantum well valence bands at higher binding energies. A gap of size 0.4 eV separates the valence and conduction bands (VB and CB, as labeled in the figure). Sb possesses the same topological $Z_2$ order as the other common topological insulators. This topological order is revealed in the present case by counting the quantum well bulk bands at two time-reversal invariant momenta, $\bar{M}$ and $\bar{\Gamma}$. Tracking the continuous quantum well bands from $\bar{\Gamma}$ to $\bar{M}$, we find that one of the Rashba surface bands near $\bar{\Gamma}$ connects to the top quantum well valence band at $\bar{M}$ while the other Rashba surface band connects to the bottom quantum well conduction band at $\bar{M}$, as indicated by the arrows in Fig.~2(a). The topology of the connection pattern is equivalent to the topological Dirac cone bridging the conduction and valence bands in the bulk case, which is the signature of the nontrivial topological $Z_2$ order. 

The overall band picture including the unoccupied bands is visualized by the results of first-principles calculations shown in Fig.~2(b), which are in excellent agreement with the ARPES data. This agreement confirms the structural quality of the films. The energies of the quantum well states depend on the film thickness.  If there existed in the film domains with different thicknesses, the quantum well bands for different thicknesses would mix up, which is not the case in our experiment. We also calculate the charge density distribution of two typical states, A and B, as indicated in Fig.~2(b). As shown in Fig.~2(c), A is a surface state and B is a quantum well state, with qualitatively different change distributions. 

Both the surface states and the quantum well states are two-dimensional states, so they are dispersionless in energy when the photon energy used for ARPES measurements is varied. The experimental results for the 20 BL Sb film around $\bar{\Gamma}$ are shown in Fig.~3. While the bands remain at the same energies, their intensities are strongly modulated. At 16-18 eV, the surface states are the dominant features. Starting from 19 eV, the intensity near the top of the valence band begins to increase. As the photon energy increases further, there appears to be an energy range of about 0.5 eV wide wherein the intensity of the states is maximized, and this resonance region moves toward higher binding energy continuously. This is very similar to the direct transition resonance in the bulk \cite{Hansen}, but here individual discrete states undergo resonances and the overall behavior is governed by a resonance envelope function akin to the bulk resonance. 

The physics of the resonance behavior is revealed by a theoretical simulation based on  first-principles calculations of the band structure. The experimental geometry is shown in Fig.~4(a). The photoexcitation process, within the so-called one-step model, involves the transition matrix elements $M_{if}=\langle\psi_{f}|\Delta H|\psi_{i}\rangle$, where the interaction Hamiltonian consists of three parts:
\begin{equation}
\Delta H\propto \textit{\textbf{A}}\cdot\nabla+\frac{1}{2}\nabla\cdot\textit{\textbf{A}}+\alpha\textit{\textbf{s}}\cdot\textit{\textbf{A}}\times\nabla V
\end{equation}
where $\textit{\textbf{A}}$ is the vector potential of the incident light. The first term corresponds to dipole transition and is the dominant contribution for bulk direct transitions governed by momentum conservation in all three dimensions. The second term arises from surface photoemission and is important only at the surface where the dielectric function $\epsilon$ and hence the vector potential $\textit{\textbf{A}}$ are discontinuous \cite{Tom}. The dielectric function of Sb in our calculation is taken from \cite{dielectric}. The discontinuity results in a $\nabla\cdot\textit{\textbf{A}}$ peaked around the surface. The third term depends on the spin $\textit{\textbf{s}}$ explicitly and stems from the Rashba interaction \cite{Rashba, Gedik} and it also peaks around the surface. For simplicity, we consider only the spectrum taken at the normal emission angle. The spin-orbit term vanishes and the matrix element becomes
\begin{equation}
M_{if}\propto\langle\psi_{f}|\textit{\textbf{A}}\cdot\nabla+\frac{1}{2}\nabla\cdot\textit{\textbf{A}}|\psi_{i}\rangle.
\end{equation}

The initial states are taken from our first-principles calculation. The final states ($\text{TRLEED}$ states) are approximated by keeping only the leading order terms \cite{TC, Caizhi}:
\begin{eqnarray*}
\psi(\textit{\textbf{r}})&=&\left[e^{ik'_{z}z}+r_{1}e^{-ik'_{z}z}+ \frac{t_{1}t_{2}r_{3}}{1-r_{2}r_{3}e^{-2\lambda d}} e^{-2\lambda d}e^{-ik'_{z}z}\right]e^{ik_{x}x+ik_{y}y},~~\text{for $z>$0 (vacuum)} \\
&=&\left[ \frac{t_{1}}{1-r_{2}r_{3}e^{-2\lambda d}}e^{ik_{z}z+\lambda z}+\frac{t_1 r_3}{1-r_{2}r_{3}e^{-2\lambda d}}e^{-ik_{z}z-\lambda z}e^{-2\lambda d}\right]e^{ik_{x}x+ik_{y}y},~~\text{for $z<$0 (film)}
\end{eqnarray*}
where $d$ is the film thickness, 1$\slash\lambda$ is the damping factor ($\lambda$ is chosen to be 12 $\text{\AA}$; the results are roughly the same for
any value between 10 and 16 $\text{\AA}$),  $k'_z$ and $k_z$ are the $z$-component of the electron wave vector in vacuum and inside the film, respectively,  and $r$ and $t$ are the reflection and transmission coefficients noted in Fig.~4(b), respectively:
\begin{equation}
k'_{z}=\sqrt{\frac{2m_e}{\hbar^2}(h\nu-E_{B}-W)-k_{x}^{2}-k_{y}^{2}},~k_{z}=\sqrt{\frac{2m_e}{\hbar^2}(h\nu-E_{B}-W+U_{0})-k_{x}^{2}-k_{y}^{2}}
\end{equation}
where $E_B$ is the binding energy, $W$ is the work function (4.5 eV), and $U_0$ is the inner potential (13.5 eV). The reflection and transmission coefficients at the surface are derived from $k'_z$ and $k_z$ according to the Stokes relations,
\begin{eqnarray*}
r_{1}&=&(k'_{z}-k_{z})/(k'_{z}+k_{z}),~ t_{1}=2k'_{z}/(k'_{z}+k_{z})\\
r_{2}&=&(k_{z}-k'_{z})/(k'_{z}+k_{z}),~ t_{1}=2k_{z}/(k'_{z}+k_{z}).
\end{eqnarray*}
The surface emission term can be simplified to 
\begin{equation}
\frac{1}{2}\nabla\cdot\textit{\textbf{A}}\approx\frac{1}{2}\left(1-\frac{1}{\epsilon}\right)A_{z}\delta(0)
\end{equation} 
The photoemission intensity for each state is given by
\begin{equation}
I_{PE}\propto \left|M_{if}\right|^2=\left|\langle\psi_{f}|\textit{\textbf{A}}\cdot\nabla|\psi_{i}\rangle+\frac{1}{2}\left(1-\frac{1}{\epsilon}\right)A_{z}\psi_{f}^{*}(0)\psi_{i}(0)\right|^2
\end{equation} 
The energy dispersion of the TRLEED final state is 
\begin{equation}
E_{f}=\frac{\hbar^{2}k_{z}^{2}}{2m_{e}}+W-U_{0}
\end{equation} 
The dipole transition conserves both momentum and energy, so the following relation must be satisfied for direct transition resonances,
\begin{equation}
h\nu=E_{f}(k_{z})-E_{i}(k_{z})=\frac{\hbar^{2}k_{z}^{2}}{2m_{e}}+W-U_{0}-E_{i}(k_{z})=\frac{\hbar^{2}k_{z}^{2}}{2m_{e}}+W-U_{0}+E_{B}(k_{z})
\end{equation} 
For photon energy in the range from 16 eV to 29 eV, the momentum $k_{z}$ is found to be $4k_{\Gamma T}-k$, where $k$ is the reduced crystal momentum along $\Gamma$-$T$. The band dispersion of the initial and final states is plotted in Fig.~4(c) and the allowed direct transitions are indicated by the vertical arrows. As the photon energy increases, the initial state of the direct transition resonance moves from $T$ to $\Gamma$ while shifting toward higher binding energies, in excellent agreement with the experiment. Because of the finite final state damping length, the momentum conservation rule is broadened, which explains the finite resonance width of about 0.5 eV as seen in Fig.~3. The experimental energy distribution curves (EDC) at normal emission for different photon energies are plotted in Fig.~4(d), where each peak corresponding to an initial state is broadened by a Lorentzian function with width 80 meV to simulate lifetime broadening. Evidently, the energy positions of the quantum well peaks do not move in energy but their intensities are modulated by the direct transition resonance. All these features are well reproduced by our simulation shown in the middle panel of Fig.~4(d).  We note that there exist some discrepancies in energy positions of the states between theory and experiment, which are not unexpected because of the limited accuracies of first-principles calculations. Over the entire range of photon energy studied, the dipole term predominates in the emission from bulk states. The surface photoemission term contributes less than 15\% to the total intensity from the bulk quantum well states. Figure~4(d) shows the theoretically computed spectra corresponding to the contribution of dipole term and the difference between the total intensity and the dipole emission. The surface photoemission term, being dependent on the weight of the wave function at the surface, is expected to be more relevant to the surface states. Indeed, it is shown in the simulated spectra that the surface state emission can be enhanced as much as 55\% by the inclusion of the surface photoemission term.

Referring to Figs.~3 and~4(d), it is evident that the resonant behavior of the surface states and the bulk states are tied together. This is expected based on the dipole transition. The perpendicular momentum of the surface states is $k_{T}$ plus an imaginary part \cite{Ashcroft}. The dipole selection rule still applies approximately; the imaginary part simply causes additional broadening. Generally, surface states can be viewed as boundary-specific solutions to the Schr$\ddot{\text{o}}$dinger equation that are connected to bulk states at the band edge by analytic continuation. For the topological surface states under consideration, their resonance behavior and in-plane dispersion relations demonstrate that the analytic connection happens in both the perpendicular and in-plane directions in $k$ space. 

In summary, we explored an experimental procedure to grow atomically uniform Sb(111) films via MBE. The Bi-$\sqrt{3}\times\sqrt{3}$ surface reconstruction based on a Si(111) substrate was proven to be an excellent choice for the smooth growth of Sb films with a precise control on thickness. The electronic band structure of the Sb films was mapped out by ARPES. The connection of the topological surface bands to the quantum well bulk bands shows explicitly the topological order of the Sb films. This discrete connection is only possible in the quantum film geometry rather than the bulk. We performed a theoretical analysis and simulation of the photoexcitation process taking into account first-principles calculated initial states and TRLEED final states. The results offer a comprehensive understanding of the observed evolution of a direct transition resonance with photon energy for both the quantum well states and the topological surface states. Surface photoemission plays a relatively a minor role in the present case, but it can be quite important in others \cite{Caizhi, Bian_SA, YHWang, Bian_Rashba}. 

\section{Acknowledgements}
This work was supported by the US Department of Energy (DOE) Grant No.~DE-FG02-07ER46383 (T.C.C.), the US National Science Foundation (NSF) Grant No.~NSF-DMR-1006492 (M.Z.H.) the National Science Foundation of China under Grant No. 11204133 (XW), the Jiangsu Province Natural Science Foundation of China under Grant No. BK2012393 (XW), and the Young Scholar Project of Nanjing University of Science and Technology (XW). T.R.C. acknowledges visiting scientist support from Princeton University. T.R.C. and H.T.J. are supported by the National Science Council, Academia Sinica, and National Tsing Hua University, Taiwan. We also thank NCHC, CINC-NTU, and NCTS, Taiwan for technical support. We thank M. Bissen and M. Severson for assistance with beamline operation at the Synchrotron Radiation Center, which was supported by the University of Wisconsin-Madison. T.M. and the beamline operations were partially supported by US NSF Grant No.~NSF-DMR-1305583.

\newpage

\begin{figure}
\centering
\includegraphics[width=16cm]{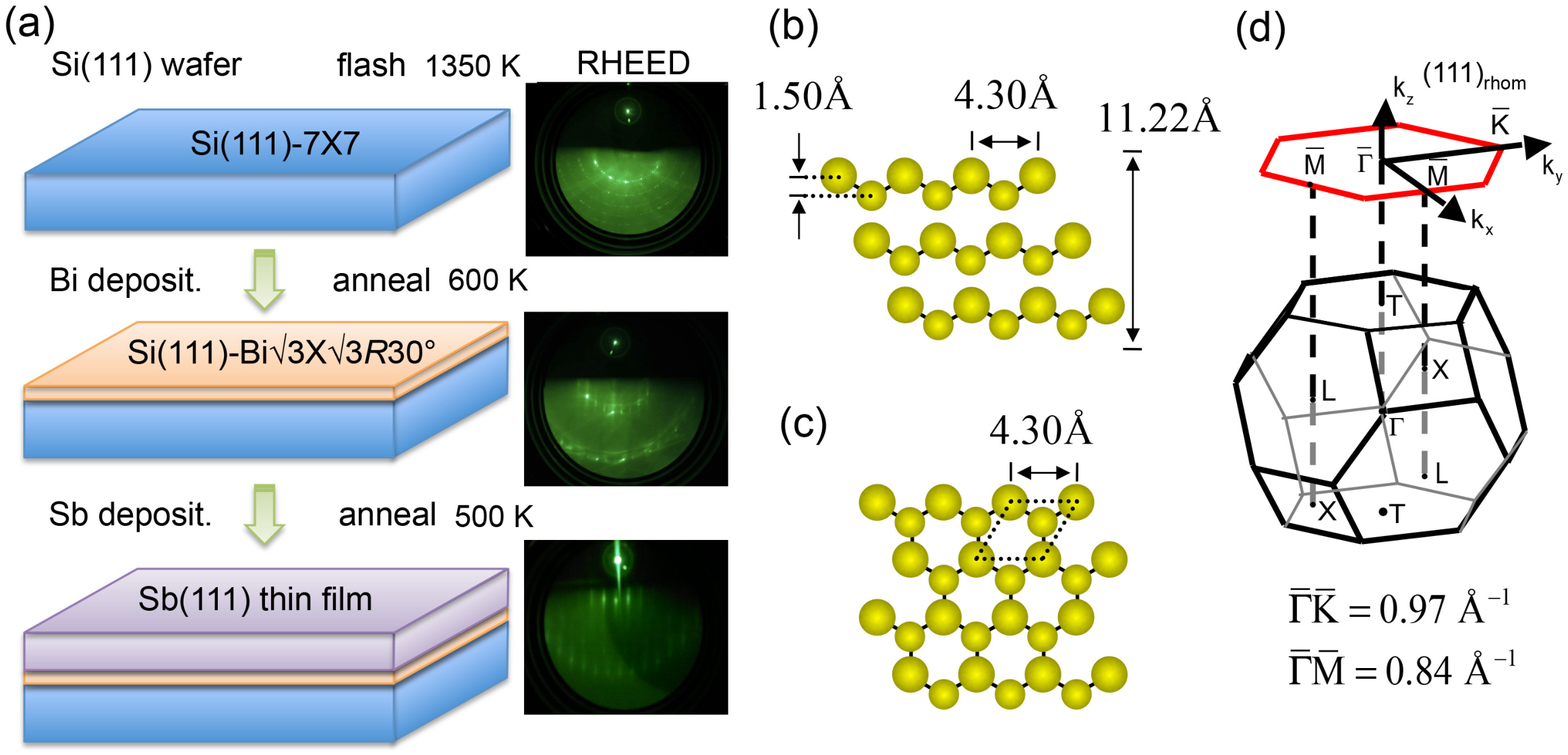}
\caption{(color online). (a) Schematic of the sample construction after each preparation step. RHEED pattern was taken after each deposition and annealing step. (b) Side view and (c) top view of  Sb(111) crystal lattice. (d) Bulk Brillouin zone and (111)$_\text{rhom}$-projected surface Brillouin zone of Sb.}
\end{figure}

\newpage

\begin{figure}
\centering
\includegraphics[width=16cm]{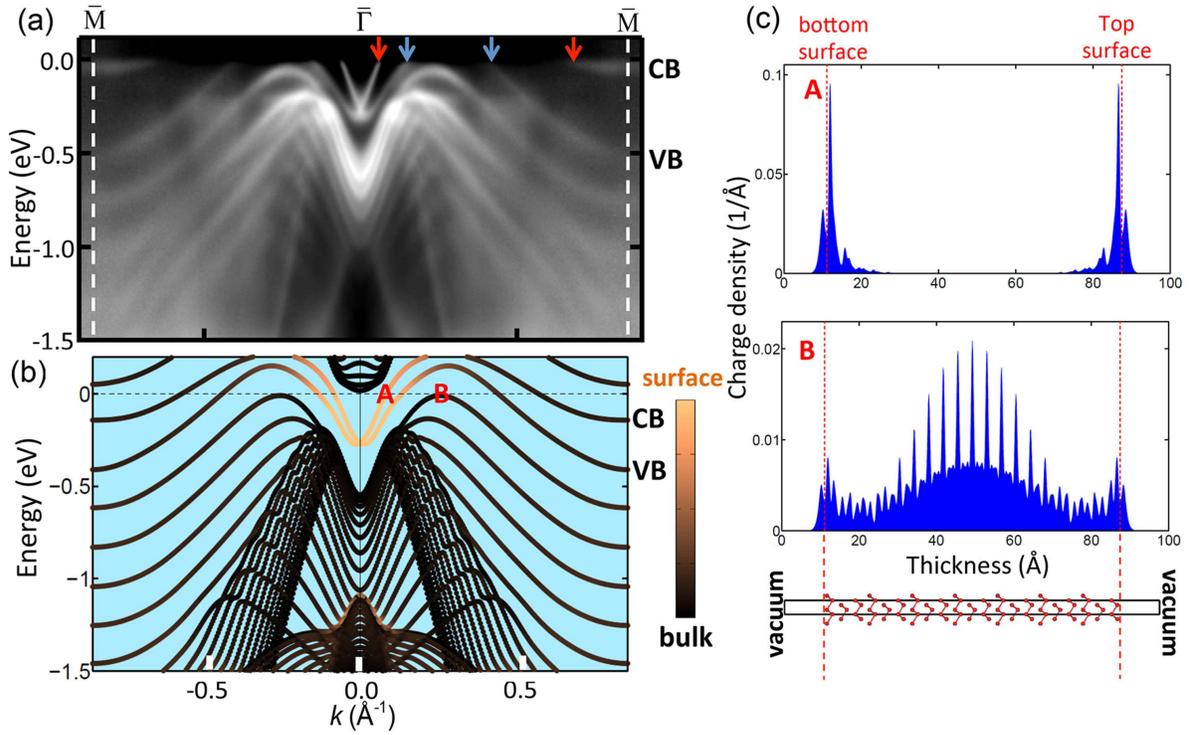}
\caption{(color online). (a) ARPES spectrum taken from a 20-BL Sb(111) film grown by MBE. The photon energy is 22 eV. The conduction (CB) and valence bands (VB) are indicated. (b) Calculated  band structure of a 20 BL Sb(111) slab. The color indicates the surface weight of the corresponding wavefunction. (c) Charge distribution of the states A and B as marked (b).} 
\end{figure}

\newpage

\begin{figure}
\centering
\includegraphics[width=16cm]{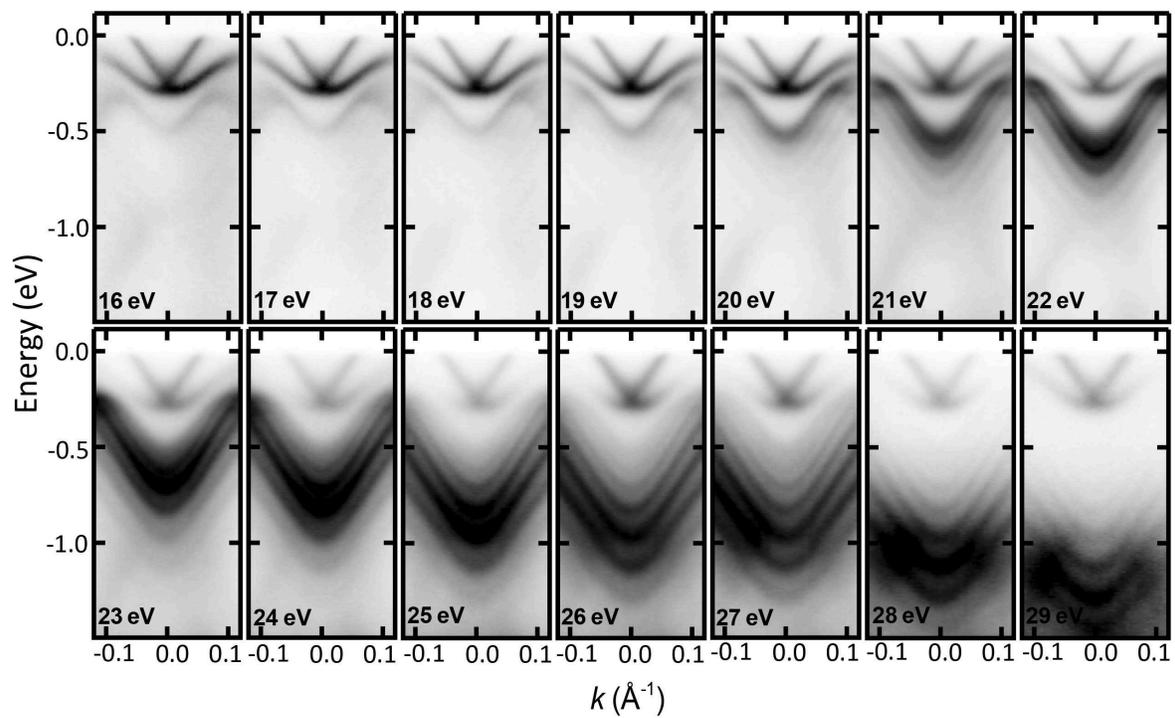}
\caption{(color online). ARPES spectra of a 20-BL Sb(111) film taken with photon energies from 16 eV to 29 eV.}
\end{figure}

\newpage

\begin{figure}
\centering
\includegraphics[width=16cm]{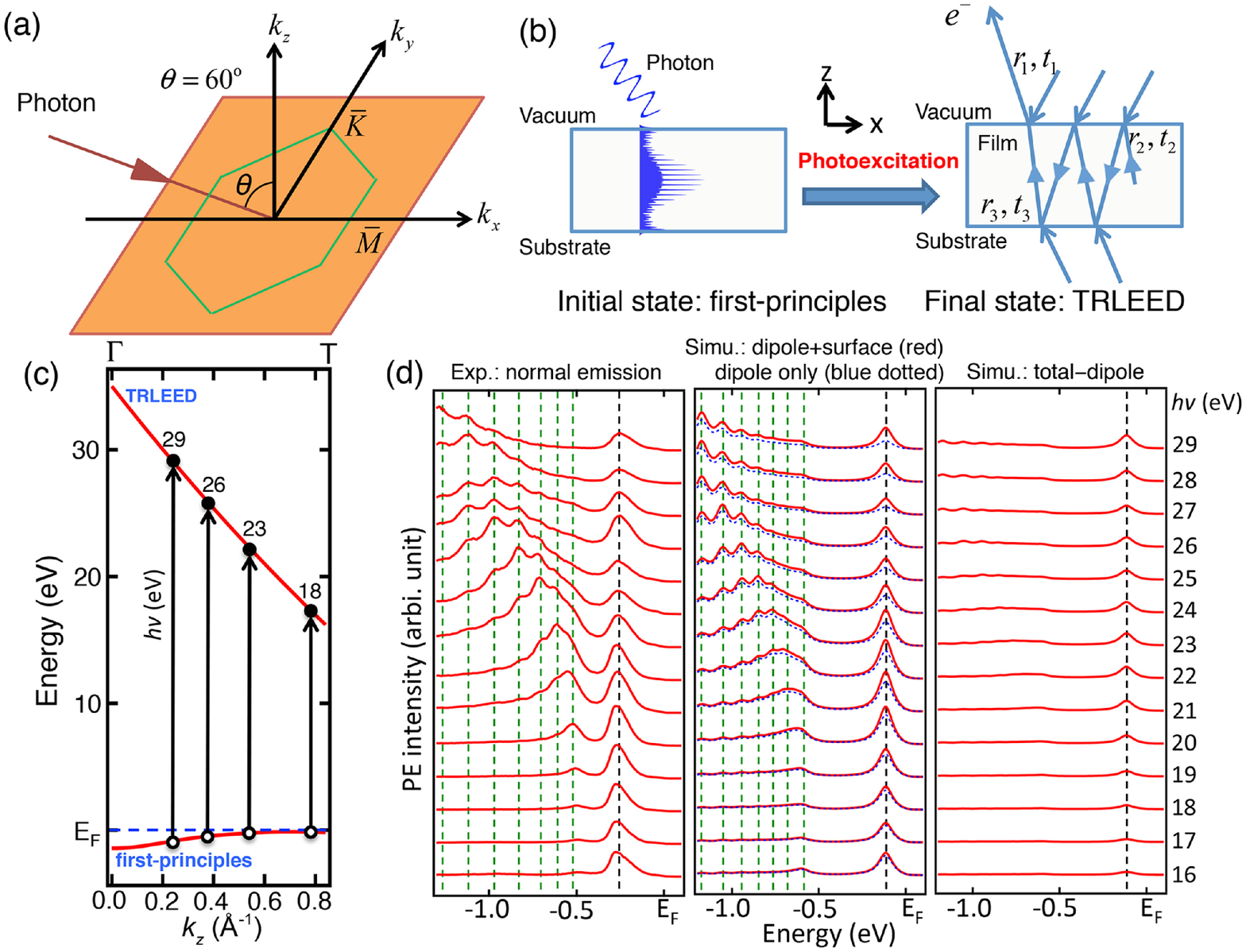}
\caption{(a) Experimental geometry of ARPES measurement. (b) Schematic of the photoexcitation process. The final state can be considered as a time-reversed low-energy-electron-diffraction (TRLEED) state. (c) Band structure of Sb along the [111]$_\text{rhom}$ direction. The direct transition resonances for photon energies of 18, 23, 26, and 29 eV are indicated by vertical arrows. (d) Energy distribution curves (EDC) for various photon energies. Left panel: ARPES experiment; middle panel: simulation results considering both dipole direct transition and surface emission (red solid lines), and dipole term only (blue dotted lines); and right panel: difference between the total intensity and the contribution from the dipole term. }
\end{figure}


\begin{thebibliography}{99}

\bibitem{Ehrlich} A. C. Ehrlich, Phys. Rev. Lett.  $\mathbf{71}$, 2300 (1993).

\bibitem{Xue} Y. Guo, Y.-F. Zhang, X.-Y. Bao, T.-Z. Han, Z. Tang, L.-X. Zhang, W.-G. Zhu, E. G. Wang, Q. Niu, Z. Q. Qiu, J.-F. Jia, Z.-X. Zhao, and Q.-K. Xue, Science $\mathbf{306}$, 1915 (2004).

\bibitem{Liu} C. Liu, H. Zhang, B. Yan, X.-L. Qi, T. Frauenheim, X. Dai, Z. Fang, and S.-C.Zhang, Phys. Rev. B $\mathbf{81}$, 041307 (2010).

\bibitem{TC} T.-C. Chiang, Surf. Sci. Rep. $\mathbf{39}$, 181 (2000).

\bibitem{LaShell} S. Lashell, B. A. MacDougall, and E. Jensen, Phys. Rev. Lett. $\mathbf{77}$, 3419 (1996).

\bibitem{Bihlmayer} L. El-Kareh, G. Bihlmayer, A. Buchter, H Bentmann, S. Bl$\ddot{ü}$gel, F. Reinert, and M. Bode, New J. Phys. $\mathbf{16}$, 045017 (2014).

\bibitem{Schouteden} K. Schouteden, P. Lievens, C. Van Haesendonck, Phys. Rev. B $\mathbf{79}$, 195409 (2009).

\bibitem{Ast} C. R. Ast and H. H$\ddot{o}$chst, Phys. Rev. Lett. $\mathbf{87}$, 177602 (2001).

\bibitem{Bian_Bi} G. Bian, T. Miller, and T.-C. Chiang, Phys. Rev. B $\mathbf{80}$, 245407 (2009).

\bibitem{Hirahara} T. Hirahara, J. Elec. Spec, and Relat. Phenom. $\mathbf{201}$, 98 (2015).

\bibitem{Hofmann} P. Hofmann, Prog. Surf. Sci. $\mathbf{81}$, 191 (2006).

\bibitem{Hasan} M. Z. Hasan, and C. L. Kane, Rev. Mod. Phys. $\mathbf{82}$, 3045 (2010).

\bibitem{Zhang} X.-L. Qi, and S.-C. Zhang, Rev. Mod. Phys. $\mathbf{83}$, 1057 (2011).

\bibitem{Hasegawa} T. Hirahara, T. Nagao, I. Matsuda, G. Bihlmayer, E.V. Chulkov, Yu.M. Koroteev, P.M. Echenique, M. Saito, and S. Hasegawa, Phys. Rev. Lett. $\mathbf{97}$, 146803 (2006).

\bibitem{Yang} Y. Liu, J. J. Paggel, M. H. Upton, T. Miller, and T.-C. Chiang, Phys. Rev. B $\mathbf{78}$, 235437 (2008).

\bibitem{Shih} D. Eom, S. Qin, M.-Y. Chou, and C. K. Shih, Phys. Rev. Lett. $\mathbf{96}$, 027005 (2006).

\bibitem{ZhangYi} Y. Zhang, K. He, C.-Z. Chang, C.-L. Song, L.-L. Wang,	X. Chen, J.-F. Jia, Z. Fang, X. Dai, W.-Y. Shan, S.-Q. Shen, Q. Niu, X.-L. Qi, S.-C. Zhang, X.-C. Ma, and Q.-K. Xue, Nat. Phys. $\mathbf{6}$, 584 (2010).

\bibitem{LiuYang} Y. Liu, G. Bian, T. Miller, M. Bissen, and T.-C. Chiang, Phys. Rev. B $\mathbf{85}$, 195442 (2012).

\bibitem{MoS2} Y. Zhang, T.-R. Chang, B. Zhou, Y.-T. Cui, H. Yan, Z. Liu, F. Schmitt, J. Lee, R. Moore, Y. Chen, H. Lin, H.-T. Jeng, S.-K. Mo, Z. Hussain, A. Bansil, and Z.-X. Shen, Nat. Nanotechnol.  $\mathbf{9}$, 111 (2014).

\bibitem{Bian_Sb} G. Bian, T. Miller, and T.-C. Chiang, Phys. Rev. Lett. $\mathbf{107}$, 036802 (2011).

\bibitem{HGH} C. Hartwigsen, S. Goedecker, and J. Hutter, Phys. Rev. B $\mathbf{58}$, 3641 (1998).

\bibitem{abinit} X. Gonze et al., Comput. Phys. Commun. $\mathbf{180}$, 2582 (2009).

\bibitem{Hansen} E. D. Hansen, T. Miller, and T.-C. Chiang, Phys. Rev. B $\mathbf{55}$, 1871 (1997).

\bibitem{Tom} T. Miller, W. E. McMahon, and T.-C. Chiang, Phys. Rev. Lett. $\mathbf{77}$, 1167 (1996).

\bibitem{dielectric} M. Cardona and D. L. Greenaway, Phys. Rev. $\mathbf{133}$, A1685 (1964).

\bibitem{Rashba} Y. A. Bychkov, and E. I. Rashba, J. Phys. C: Solid State Phys. $\mathbf{17}$, 6039 (1984).

\bibitem{Gedik} Y. H. Wang, D. Hsieh, D. Pilon, L. Fu, D. R. Gardner, Y. S. Lee, and N. Gedik, Phys. Rev. Lett. $\mathbf{107}$, 207602 (2011).

\bibitem{Ashcroft} N. W. Ashcroft and N. D. Mermin, Solid State Physics (Thomson Learning, Inc. 1976).

\bibitem{Caizhi} C.-Z. Xu, Y. Liu, R. Yukawa, L.-X. Zhang, I. Matsuda, T. Miller, and T.-C. Chiang, Phys. Rev. Lett. $\mathbf{115}$, 016801 (2015).

\bibitem{Bian_SA} G. Bian, L. Zhang, Y. Liu, T. Miller, and T.-C. Chiang, Phys. Rev. Lett. $\mathbf{108}$, 186403 (2012).

\bibitem{YHWang} Y. Wang and N. Gedik, Phys. Status Solidi RRL 7, $\mathbf{1}$-$\mathbf{2}$, 64 (2013).

\bibitem{Bian_Rashba} G. Bian, T. Miller, and T.-C. Chiang, J. Elec. Spec, and Relat. Phenom. $\mathbf{201}$, 36 (2015).

\end{thebibliography}
\end{document}